\documentclass[10pt]{article}
\usepackage[dvips]{graphicx}

\title{Structural instability 
       of two- and three-dimensional pyrochlore spin lattices
       in high magnetic fields}
\author{Oleg Derzhko$^{1,2,3}$
        and
        Johannes Richter$^2$\\
\small{$^1$Institute for Condensed Matter Physics,
       National Academy of Sciences of Ukraine,}\\
\small{1 Svientsitskii Street, L'viv-11, 79011, Ukraine}\\
\small{$^2$Institut f\"{u}r Theoretische Physik,
       Universit\"{a}t Magdeburg,}\\
\small{P.O. Box 4120, D-39016 Magdeburg, Germany}\\
\small{$^3$National University ``Lvivska Politechnika'',}\\
\small{12 S.~Bandera Street, L'viv, 79013, Ukraine}}

\date{\today}

\begin{document}

\renewcommand\baselinestretch {1.1}
\large\normalsize

\maketitle

\begin{abstract}
We consider the quantum spin-$s$ $XXZ$ Heisenberg antiferromagnet 
on the two- and three-dimensional pyrochlore lattices
and examine a spin-Peierls mechanism
of lowering the total energy by a lattice distortion
in a high magnetic field.
For the exact eigenstates
consisting of several independent localized magnons
in a ferromagnetic environment
we show the existence of  a spin-Peierls instability
by rigorous analytical calculations.
In addition we report 
on exact diagonalization data for finite two-dimensional pyrochlore lattices
up to $N=64$ sites.
We discuss a scenario
of the field-tuned changes in the ground-state properties
of the frustrated spin lattices
due to the coupling between spin and lattice degrees of freedom.
\end{abstract}

\vspace{2mm}

\noindent
{\bf PACS number(s):}
75.10.Jm, 75.45.+j

\vspace{2mm}


\vspace{5mm}

\renewcommand\baselinestretch {1.43}
\large\normalsize

\section{Introduction. Independent localized-magnon states}

Antiferromagnetically interacting Heisenberg spins
on geometrically frustrated lattices
have attracted much attention of physicists during last years.
Frustrating interactions may lead to a rich phase diagram 
both on the classical and quantum levels
\cite{01,02,03,04}.
In the presence of an external magnetic field
frustrated spin systems exhibit
a number of specific properties.
In particular,
plateaus and jumps can be observed
in the zero-temperature magnetization curve
for such models.
Another intriguing property
is a magnetic-field induced spin-Peierls effect in frustrated systems
to be discussed below.

A particular case of geometrically frustrated lattices
is the pyrochlore lattice
(i.e. a network of corner-sharing tetrahedra)
which is especially attractive 
since it has many experimental realizations.
The two-dimensional (2D) version of the pyrochlore network,
also known as the planar pyrochlore, 
the checkerboard lattice, 
or the square lattice with crossings,
has the same local coordination as in the three-dimensional (3D) case,
however, it is easier for numerical studies.
The studies of this simpler model 
can be helpful in understanding of its 3D analog
\cite{05,06,07,08}.
 
There is a number of experimental and theoretical papers
discussing the effects of magnetoelastic couplings
in geometrically frustrated antiferromagnets.
Because of frustration,
such systems can have a huge degeneracy of the ground state
which prevents the system from finding a unique ground state 
as temperature decreases.
Any perturbation can have a dramatic effect 
selecting one ground state over another.
Different perturbations which can lift the ground-state degeneracy
have been discussed in the literature,
and a magnetoelastic coupling is among them.
This coupling allows the spin lattice to distort
thus relieving the magnetic frustration 
and lowering the total energy of the system.
Much attention in recent years has been paid to 
such spin-Peierls-like phase transitions
in the pyrochlore antiferromagnets from the experimental as well as 
from the theoretical side.
Thus, inelastic magnetic neutron scattering on the 
3D pyrochlore antiferromagnet ZnCr$_2$O$_4$
(the spin-$3/2$ Cr$^{3+}$ ions 
form the 3D pyrochlore lattice)
revealed that a lattice distortion can lower the energy
driving the spin system into an ordered phase \cite{09}.
Further probes of a lattice distortion in ZnCr$_2$O$_4$ 
using neutron scattering and infrared spectroscopy have been reported 
in Refs. \cite{10,11}.
NMR investigation
of 
Y$_2$Mo$_2$O$_7$
(the spin-$1$ Mo$^{4+}$ ions 
form the 3D pyrochlore lattice)
gives evidence
for discrete lattice distortions
which reduce the energy of the system
\cite{12}.
The results of a similar $\mu$SR study for this compound 
have been reported recently in Ref. \cite{13}.
A lifting of a macroscopic ground-state degeneracy
of frustrated magnets
through a coupling between spin and lattice degrees of freedom
in pyrochlore antiferromagnets
was studied theoretically in Refs. \cite{14,15,16}
and the relation of these theoretical studies 
to pyrochlore compounds was discussed.
More recently,
inspired from the experimental data 
on ZnCr$_2$O$_4$
\cite{10}
the authors of Ref. \cite{17}
have examined 
a model 
of lattice-coupled antiferromagnetic classical spins 
on the pyrochlore lattice.
They conclude
that a particular pattern 
-- a hexagon contraction 
-- may arise owing to a lattice distortion
which reduces the energy.  
While the above mentioned studies refer to zero-field case, 
Penc and coworkers \cite{18,19}
have considered the magnetoelastic coupling at finite magnetic fields 
and have shown
that the coupling between classical spin and lattice degrees of freedom 
in a pyrochlore antiferromagnet 
stabilizes a half-magnetization plateau
which occurs at intermediate fields.
This conclusion agrees with the high-field measurements 
at up to 47 T 
for the $s=3/2$ Heisenberg pyrochlore antiferromagnet CdCr$_2$O$_4$ 
\cite{20}.

In what follows,
we want to discuss the impact of localized-magnon states 
\cite{21,22,23}
on the stability 
of the 2D and 3D pyrochlore lattices 
against distortions. 
In contrast to the above mentioned zero-field studies
on the effects of magnetoelastic couplings
in the pyrochlore lattice,
high magnetic fields are essential in our consideration
since only in a high magnetic field
the independent localized magnons
may become relevant for the ground-state properties
of the considered systems.
The present study extends our earlier results 
referring to the square-kagom\'{e} and kagom\'{e} lattices
\cite{24,25}. 

\begin{figure}
\begin{center}
\includegraphics[clip=on,width=9.0cm,angle=0]{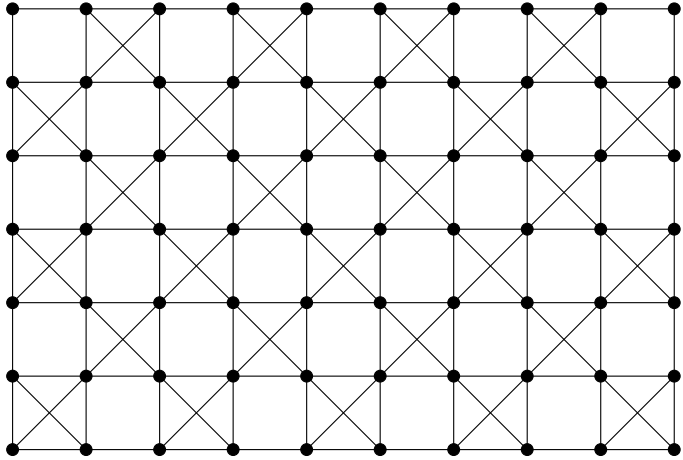}
\caption
{The undistorted two-dimensional pyrochlore lattice. 
All bonds have the same strength $J$.
In what follows 
we use as shorthands the notations 
``empty squares'' 
(squares without bond-crossings) 
and ``crossed squares'' 
(squares with bond-crossings 
which are, in fact, projected tetrahedra).}
\label{fig01}
\end{center}
\end{figure}

\begin{figure}
\begin{center}
\includegraphics[clip=on,width=9.0cm,angle=0]{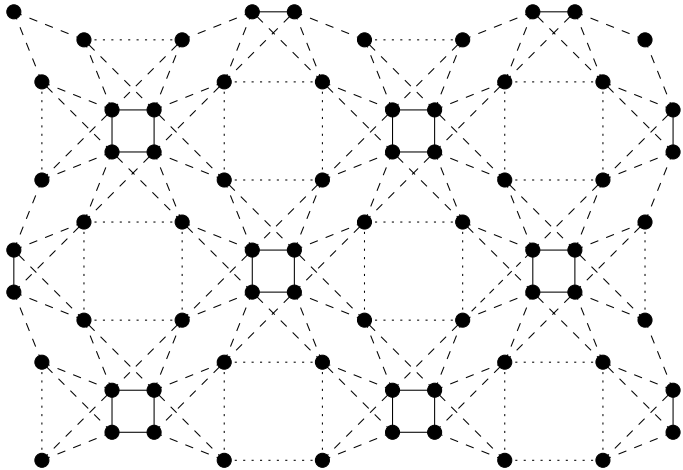}
\caption
{The distorted two-dimensional pyrochlore lattice. 
The solid and the dotted bonds have the strengths 
$J\left(1+\sqrt{2}\delta\right)$
and
$J\left(1-\sqrt{2}\delta\right)$, 
respectively,
the dashed bonds have the strength $J$ 
(with an accuracy up to $\delta$).
Note
that the lattice volume remains unchanged after such a distortion.}
\label{fig02}
\end{center}
\end{figure}

\begin{figure}
\begin{center}
\includegraphics[clip=on,width=9.0cm,angle=0]{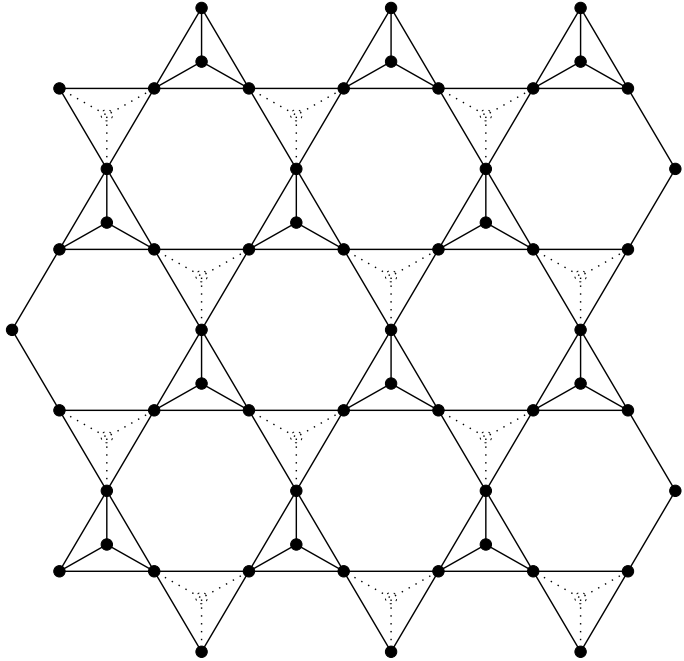}
\caption
{The $\{111\}$ slice 
of the undistorted three-dimensional pyrochlore lattice.
All bonds have the same strength $J$.}
\label{fig03}
\end{center}
\end{figure}

\begin{figure}
\begin{center}
\includegraphics[clip=on,width=9.0cm,angle=0]{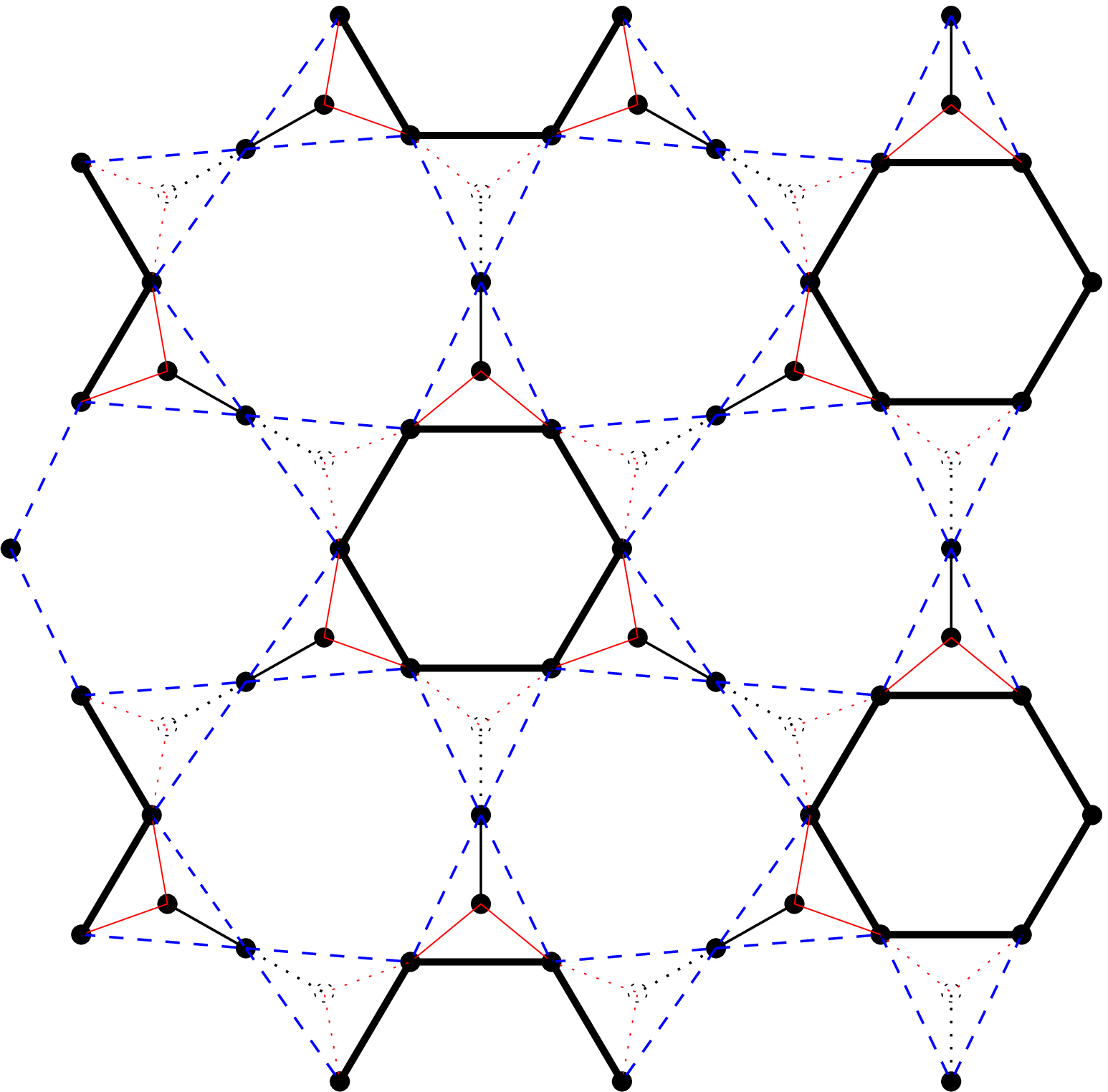}
\caption
{(color online). 
The distorted $\{111\}$ slice 
of the three-dimensional pyrochlore lattice.
The thick solid bonds 
(thick black)
represent the shrinked hexagons and have the strength $J(1+\delta)$,
the thin solid and dotted bonds 
(red)
have the strength $J\left(1-\delta/\left(2\sqrt{3}\right)\right)$,
the dashed bonds 
(blue)
have the strength $J\left(1-\delta/2\right)$.
The normal solid and dotted bonds 
(normal black)
remain as in the undistorted lattice 
having the strength $J$.
Note
that the lattice volume remains unchanged after such a distortion.}
\label{fig04}
\end{center}
\end{figure}

To be specific,
we consider two geometrically frustrated lattices,
namely,
the 2D pyrochlore lattice
(Figs. \ref{fig01}, \ref{fig02})
and
the 3D pyrochlore lattice 
(Figs. \ref{fig03}, \ref{fig04}).
The Hamiltonian
of $N$ quantum spins of length $s$
reads
\begin{eqnarray}
\label{01}
H=
\sum_{(nm)}J_{nm}\left(\frac{1}{2}\left(s_n^+s_m^-+s_n^-s_m^+\right)
+\Delta s_n^zs_m^z\right)
-hS^z.
\end{eqnarray}
Here
the sum runs over the bonds (edges)
which connect the neighboring sites (vertices)
on the spin lattice under consideration,
$J_{nm}>0$ are the antiferromagnetic exchange constants
between the sites $n$ and $m$,
$\Delta\ge 0$ is the exchange interaction anisotropy parameter,
$h$ is the external magnetic field,
and
$S^z=\sum_ns_n^z$ is the $z$-component of the total spin.
We assume that all bonds in the lattice
without distortion
have the same length
and hence all exchange constants have the same value $J$.

We begin with a brief illustration of 
the independent localized-magnon states in the considered systems
(the localized magnons 
in the uniform 2D and 3D pyrochlore lattices 
were also considered in Refs. \cite{23} and \cite{22}, respectively;
for a review on localized magnons see Refs. \cite{26,27}). 
The localized-magnon states are excitations above the fully polarized
ferromagnetic state, which is the ground state of the antiferromagnet in
magnetic fields exceeding the saturation field $h_1$. 
The localized magnons live on restricted areas of the
lattice, e.g. on the empty squares of the 2D pyrochlore lattice 
or on the hexagons of 3D pyrochlore lattice. They represent exact
eigenstates of the frustrated spin lattice and become ground states in high
magnetic fields.    

In the 2D pyrochlore lattice without distortion 
the energy of the eigenstate with one magnon
localized on an empty square 
is 
\begin{eqnarray}
\label{02}
E_1=-2sJ + 2s\left(2s-1\right)J\Delta
+ 4s\left(4s-1\right)J\Delta 
+ \left(3N-20\right)s^2J\Delta.
\end{eqnarray}
Here 
the first and the second terms represent the contribution 
from the bonds along the square,
the third term represents the contribution 
from the bonds which connect the square with the environment,
and the fourth term represents the contribution 
from the ferromagnetic environment.
For $n$ independent localized magnons 
we have instead of (\ref{02})
\begin{eqnarray}
\label{03}
E_n=3Ns^2J\Delta - n\epsilon_1,
\;\;\;
\epsilon_1=2sJ\left(1+3\Delta\right)
\end{eqnarray}
and $n$ may vary from 1 to $n_{\max}=N/8$ \cite{23}.
The maximal value of the filling number $n_{\max}$ 
corresponds 
to a dense magnon crystal 
in which magnons sit on every forth empty square.
This magnon crystal state is four-fold degenerate 
and it breaks spontaneously 
the translational symmetry of the lattice.

Let us pass to the 3D pyrochlore lattice 
without distortions. 
This lattice can be composed from alternating
kagom\'{e}-like and triangular planes
stacked along the $[111]$ direction.
The localized magnons can exist in a kagom\'{e}-like plane
(i.e.
the $\{111\}$ slice of the 3D pyrochlore lattice
considered in Ref. \cite{28})
and are trapped on hexagons.
The $\{111\}$ slice 
of the undistorted 3D pyrochlore lattice 
is shown in Fig. \ref{fig03}.
The eigenstate with one magnon trapped on a hexagon has the energy
given by Eq. (\ref{02})
(however,
with the following contributions of different bonds:
$-2sJ + 2s\left(3s-1\right)J\Delta$ (along the edges of the hexagon),
$4s\left(6s-1\right)J\Delta$ (along the edges 
connecting the hexagon with the environment),
$\left(3N-30\right)s^2J\Delta$ (ferromagnetic environment)).
For $n$ independent localized magnons 
we again arrive at Eq. (\ref{03}),
however,
$n$ may vary from 1 to $n_{\max}=N/12$ \cite{22}.
The magnon crystal state corresponds 
to maximum filling $n_{\max}$.
In this magnon crystal state 
every third hexagon in the $\{111\}$ slice is occupied by magnon
producing thus three-fold degeneracy within a slice
(as for the kagom\'{e} lattice). 
Since these three different crystal states
can be chosen independently in each slice,
the degeneracy of the magnon crystal 
of the whole 3D pyrochlore lattice 
is much higher and scales at least 
as
$3^{L}$ 
where 
${L}\sim N^{1/3}$
is the number of the $\{111\}$ slices 
in the 3D pyrochlore lattice.

Each magnon decreases $S^z$ by 1 
and the localized-magnon state with $n$ independent localized magnons
has $S^z=Ns-n$.
These states have the lowest energies 
in the corresponding sectors of $S^z$ \cite{21,29}.
Hence,
Eq. (\ref{03}) 
with $n=Ns-S^z$ gives $E_{\min}(S^z)$
for the 2D and 3D 
pyrochlore antiferromagnets 
when $S^z=Ns,Ns-1,\ldots,Ns-n_{\max}$.

In the presence of an external field 
we must add to the energy (\ref{03}) 
the Zeeman term 
$-hS^z=-h\left(Ns-n\right)$.
Then the ground state belongs to an $S^z$ determined by the condition
$h=\partial E(S^z)/\partial S^z$.
At the saturation field $h_1=\epsilon_1$ 
the energy of all localized-magnon states 
with $n=1,\ldots,n_{\max}$ 
is the same 
and the zero-temperature magnetization $S^z$ 
jumps between the saturation value $Ns$ 
and the value $Ns-n_{\max}$.

On the basis of general arguments \cite{30,31} 
(see also Refs. \cite{04,24,27})
we expect 
that the magnon crystal state 
has gapped excitations 
and hence the system exhibits 
a magnetization plateau at $S^z=Ns-n_{\max}$
between $h_2=h_1-\Delta h$ and $h_1$.
Numerical data support this expectation.
Thus,
a finite-size analysis of the data for the plateau width 
obtained for the spin-$1/2$ isotropic Heisenberg antiferromagnet on finite 2D
pyrochlore lattices 
with $N=24,32,40,48,64$ 
yields for 
$\Delta h$ about $0.2J$.
Since we know 
e.g. 
from the triangular lattice \cite{honecker,chubukov} 
and 
the square-kagom\'{e} and kagom\'{e} lattices \cite{24,25}, 
that the plateau width depends on the spin
quantum number $s$ we have also calculated the plateau width $\Delta h$ 
for the spin-$s$ isotropic Heisenberg system on the 2D
pyrochlore lattice with $N=24$ sites 
for the values of $s=1/2,\;1,\;3/2,\;2,\;5/2$.
We find that  
$\Delta h$ scales excellently as $\sqrt{s}$ leading  
to a relative plateau width $\Delta h/h_1 \sim 1/\sqrt{s}$ 
vanishing in the classical limit  $s \to \infty$. 

Finally,
we want to underline that 
the ground state of both pyrochlore quantum antiferromagnets 
exhibits a huge degeneracy 
at the saturation field 
owing to the localized-magnon states.
This happens because a certain trapping cell of the lattice 
can be either occupied by a magnon or not
whereas the energies of the states with different number of magnons 
are the same.
As a result, 
the degeneracy ${\cal{W}}$, 
that is the number of ways to place magnons on the lattice,
grows exponentially with $N$ 
with a crude estimate
${\cal{W}}>2^{n_{\max}}$ \cite{04}. 
This estimate of the ground-state degeneracy
can be improved after mapping the lattice with localized magnons 
onto an auxiliary lattice occupied by hard-core objects
\cite{32,33,34,27}.

\section{Lattice instability in high magnetic fields}

We want to examine the lattice stability 
with respect to a spin-Peierls mechanism
in high magnetic fields.
For that 
we consider a small lattice deformation 
which does not violate the conditions 
for the existence of localized-magnon ground states,
and analyze the change of the total energy 
which consists of the magnetic and the elastic parts.
We search for a maximal gain of the magnetic energy 
due to lattice deformation.
For the magnon crystal state the  Eqs. (\ref{02}), (\ref{03}) give a hint 
how such a deformation may be constructed.
Actually,
for the magnon crystal state 
the nearest-neighbor spin correlation functions 
are distributed inhomogeneously.
Along the empty square or hexagon hosting a localized magnon 
the transverse correlations 
$\langle s_i^xs_{i+1}^x\rangle+\langle s_i^ys_{i+1}^y\rangle$ 
are negative, 
but all other nearest-neighbor correlations are positive.
Therefore,
one may expect a maximal gain of magnetic energy  
by increasing the (antiferromagnetic) bonds on the empty square or hexagon 
which hosts localized magnon
and 
decreasing the (antiferromagnetic) bonds on the attaching triangles. 
The corresponding deformations 
for the 2D and 3D pyrochlore lattices 
are discussed below.
The changes of the magnetic energy due to such deformations 
are calculated analytically 
for the values  of $S^z$ corresponding to the magnon crystal. Furthermore we
present
numerical data for lower values of $S^z$
in the 2D case.

\subsection{Two-dimensional pyrochlore spin lattice}

A pattern of distortions which uses 
the inhomogeneous distribution of nearest-neighbor spin correlations 
in an optimal way 
and
does not violate
the conditions of existence of the magnon crystal state 
is shown in Fig. \ref{fig02}.
This lattice deformation 
corresponds to the $T_2$ vibration mode of isolated tetrahedron
(see Ref. \cite{16}),
i.e.
in every tetrahedron a pair 
of opposite to each other bonds  
(belonging to empty squares in Fig. \ref{fig01})
are stretched and contracted by the same amount.
Introducing the parameter $\delta$ 
which is proportional to the displacement of the sites
we find the following changes of the exchange interactions
(with an accuracy up to linear terms in $\delta$)
\begin{eqnarray}
\label{04}
J\to J\left(1\pm\sqrt{2}\delta\right),
\;\;\;
J\to J.
\end{eqnarray}
The stronger (weaker) bonds are shown in Fig. \ref{02} 
by solid (dotted) lines;
the bonds having unchanged strength 
are shown in Fig. \ref{02}
by dashed lines.
As a result,
the magnetic energy of the distorted magnon crystal state 
\begin{eqnarray}
\label{05}
E_{n_{\max}}(\delta)
=
3Ns^2J\Delta-n_{\max}\left(\epsilon_1+2\sqrt{2}s(1+\Delta)J\delta\right)
\end{eqnarray}
decreases proportionally to $\delta$.
The deformation of the 24 bonds
forming one cell 
belonging to a localized-magnon area
increases the elastic energy
by $8\gamma(\sqrt{2}\delta)^2$
where the parameter $\gamma$ 
is proportional to the elastic constant of the lattice. 
As a result,
for the distorted magnon crystal state 
the elastic energy 
$16\gamma\delta^2n_{\max}$
increases proportionally to $\delta^2$.
Combining the magnetic and the elastic energies 
of the deformed lattices
we find that the total energy 
for the magnon crystal state has a minimum 
at $\delta=\delta^\star=(\sqrt{2}/16)s(1+\Delta)J/\gamma\ne 0$
that explicitly demonstrates the lattice instability.

So far our results are rigorous.
Now the question arises 
whether the lattice distortion shown in Fig. \ref{02} 
remains energetically favorable for $S^z<Ns-n_{\max}$.
To discuss this question 
we perform exact diagonalization for finite lattices 
using J.~Schulenburg's {\it spinpack} \cite{spinpack}.
Assuming that the lowest magnetic energy 
for a given value of $S^z$
depends on $\delta$ like
$ E_{\min}(\delta)
=E_{\min}(0)+ A\delta^p$
and taking $\delta$ of the order $10^{-5} \ldots 10^{-4}$
we estimate the exponent $p$ numerically. 
If $p$ is less than 2 (and $A<0$) 
the lattice deformation for the considered finite lattice is favorable,
otherwise it is not.
We have calculated the exponent $p$ 
for the $s=1/2$ isotropic ($\Delta=1$) Heisenberg system 
on finite lattices of $N=32$, $40$ and $48$ sites 
in the two sectors of $S^z$
just below the sector of the magnon crystal state, 
i.e. 
for
$S^z=Ns-n_{\max}-1$ 
and 
$S^z=Ns-n_{\max}-2$. 
Note that for $N=48$ 
with $S^z = Ns-n_{\max}-2 = 16$ 
the lowest eigenvalues of a Hamiltonian matrix of size 
$6 \cdot 10^7 \times 6 \cdot 10^7$ 
must be calculated.
We obtain a gain in magnetic energy 
(i.e. $A < 0$) 
in all cases. 
In the sector  
$S^z=\frac{1}{2}N-n_{\max}-1$
we obtain 
$p=1.0,\;2.0,\;2.0$
for
$N=32,\;40,\;48$,
respectively.
The ``outrider'' for $N=32$
may be attributed to finite-size effects,
indeed
we find $p=2.0$ in the next sectors below $S^z=\frac{1}{2}N-n_{\max}-1$. 
For $N=40$ and $N=48$ the exponent remains $p=2.0$ in the sector 
$S^z=\frac{1}{2}N-n_{\max}-2$. 

Thus, 
our analytical calculations demonstrate 
a spin-Peierls lattice instability 
of the 2D pyrochlore antiferromagnet 
for the lowest energy states 
in the sectors of $S^z=Ns-n_{\max},\ldots,Ns-1$.
The lattice deformation is not favorable 
in the fully polarized state when $S^z=Ns$.
Our numerical calculations give evidence 
that it also disappears when $S^z<Ns-n_{\max}$.

Our findings imply 
that the 2D pyrochlore quantum antiferromagnet
at zero temperature
being placed in a high magnetic field $h>h_1$
becomes distorted 
when the applied field is lowered to the 
saturation field $h_1=\epsilon_1$,
and the lattice deformation disappears 
when the field becomes smaller 
than the field at the left endpoint of the plateau $h_2$.
It should be noted that a hysteresis phenomenon  
in the vicinity of the saturation field is expected:
Starting from a large magnetic field $h>h_1$ and then 
decreasing the field the distortion sets in at $h_1=\epsilon_1$.
On the other hand, starting from a field below the saturation field and then 
 increasing the field the distorted magnon crystal state survives 
until the saturation field 
of the distorted lattice
$h_1(\delta^{\star})=\epsilon_1+(1/8)s^2(1+\Delta)^2J^2/\gamma$.

\subsection{Three-dimensional pyrochlore spin lattice}

We proceed along the same lines 
switching to the 3D pyrochlore lattice.
A lattice distortion 
which does not violate the conditions 
for the existence of localized magnons 
may be constructed 
by appropriate deformations 
within the kagom\'{e} planes of the $\{111\}$ slices.
Such a pattern of distortions is illustrated in Fig. \ref{fig04}.
Note that  the triangular planes 
and the distance between the triangular and the kagom\'{e} planes
remain unchanged.
The shrinked
hexagons form a  $\sqrt{3}\times\sqrt{3}$ structure,  
 all other sites of the $\{111\}$ slice not
belonging to shrinked hexagons do not change their positions 
(compare Figs. \ref{fig03} and \ref{fig04}).
The described deformation is similar 
but not identical to the one 
considered in Ref. \cite{17}.
The deformation discussed in Ref. \cite{17} 
assumes that each hexagon contracts towards its center of mass 
and this distortion does not support the localized magnons.
Denoting by $\delta$ the parameter 
which is proportional to the displacement of the mentioned sites
we arrive at the following change of the exchange interactions 
(with an accuracy up to linear terms in $\delta$)
\begin{eqnarray}
\label{07}
J\to J\left(1+\delta\right),
\;\;\;
J\to J\left(1-\frac{1}{2\sqrt{3}}\delta\right),
\;\;\;
J\to J\left(1-\frac{1}{2}\delta\right)
\end{eqnarray}
along the edges of the contracted hexagons 
(thick black lines in Fig. \ref{fig04}),
along the out-of-plane triangles attached to the contracted hexagons
(thin solid and dotted red lines in Fig. \ref{fig04}),
along the in-plane triangles attached to the contracted hexagons
(dashed blue lines in Fig. \ref{fig04}),
respectively.
The exchange interaction along the bonds 
which are not connected to the contracted hexagons
(normal solid and dotted black lines in Fig. \ref{fig04})
remains unchanged and has the strength $J$.
The magnetic energy 
of the distorted magnon crystal state
then reads
\begin{eqnarray}
\label{08}
E_{n_{\max}}(\delta)
=
3Ns^2J\Delta
-n_{\max}\left(\epsilon_1
+s\left(2+\frac{6s+\sqrt{3}-1}{\sqrt{3}}\Delta\right)J\delta\right),
\end{eqnarray}
and the corresponding elastic energy is
$10\gamma\delta^2n_{\max}$.
The total energy achieves a minimum
at 
$\delta=\delta^{\star}
=(s/20)(2+(6s+\sqrt{3}-1)\Delta/\sqrt{3})J/\gamma\ne 0$
demonstrating the lattice instability.
Again this lattice deformation is accompanied 
by hysteresis phenomenon
with the saturation field of the distorted lattice
$h_1(\delta^{\star})
=\epsilon_1
+
(1/40)s^2(2+(6s+\sqrt{3}-1)\Delta/\sqrt{3})^2J^2/\gamma$.

Due to the higher dimensionality   
the finite-size effects for lattice sizes accessible with exact
diagonalization  
are much more important
for the 3D pyrochlore lattice 
and there are no reliable numerical data 
for the sectors below $S^z=Ns-n_{\max}$.
Therefore,
the question about the stability 
of the lattice against the considered deformations  
for magnetic fields below the plateau 
belonging to the magnon crystal state  
remains open for the 3D case.

\section{Conclusions}

In the present paper
we reported a spin-Peierls instability
in high magnetic fields 
of the quantum spin-$s$ $XXZ$ Heisenberg antiferromagnet
on the 2D and 3D 
pyrochlore lattices
in adiabatic treatment.
Since both lattices
support localized-magnon states 
which are the lowest energy states in the corresponding sectors of $S^z$
an instability of the uniform lattice near the saturation field 
has been proved by rigorous analytical calculations.
Results for smaller fields have been obtained 
by exact diagonalization.
Although 
we cannot exclude that other distortions  
(different from the ones shown in Fig. \ref{fig02} and Fig. \ref{fig04})
cannot come into the play,
we have given arguments 
that the considered deformations are the most favorable ones
taking  advantage of localized magnons in an optimal way
and
we have shown explicitly 
that the undistorted lattice is unfavorable 
in the ground state with $S^z=Ns-n_{\max},\ldots,Ns-1$.

It is interesting to note
that while the lattice deformation 
for the 2D pyrochlore lattice 
(Fig. \ref{fig02})
occurs simultaneously over all lattice 
and hence lifts the degeneracy of the ground state 
at the saturation field,
the lattice deformation 
for the 3D lattice 
(Fig. \ref{fig04})
is composed of local deformations 
and does not necessarily reduce the ground-state degeneracy at the saturation field.

From experimental point of view,
it seems to be important 
to have a magnetization plateau 
where the deformation occurs,
since that gives the possibility 
to observe the effect in a finite range of $h$.
Observation of  hysteresis 
in the magnetization and the deformation 
of pyrochlore lattices 
in the vicinity of the saturation field 
can be an experimental sign of the effect discussed theoretically in this paper.
For  experimental investigations of antiferromagnets near saturation field
$h_1$
one needs materials with comparably small exchange interaction $J$
and small value of spin $s$  
to be able to reach the high-field regime.
Materials with small spin quantum number $s$ are more appropriate, 
also because the  plateau near saturation vanishes 
with increasing  $s$.
We mention,
that in our analysis we dealt with ideal pyrochlore geometry
whereas in real compounds deviations may be expected.
However, 
there are arguments that the effects of the localized magnons 
survive under slight deviations from the ideal geometry \cite{32}.

There are now several magnetic materials 
which can be modeled by the 3D pyrochlore antiferromagnet
(for instance, 
spin-$1$ systems such as 
ZnV$_2$O$_4$,
MgV$_2$O$_4$,
Y$_2$Mo$_2$O$_7$, 
spin-$3/2$ systems such as
ZnCr$_2$O$_4$, 
CdCr$_2$O$_4$,
HgCr$_2$O$_4$;
most of them with exchange constants of about $10^2$ K) 
and there is some
hope that certain features connected with the above presented scenario 
of a lattice instability driven by a magnetic field
may be observed in antiferromagnetic compounds  with pyrochlore structure.

\vspace{5mm}

{\bf Acknowledgments:}
The present study was supported by the DFG
(project 436 UKR 17/17/03).
O.~D. acknowledges the kind hospitality of the Magdeburg University
in the autumn of 2004
and
in the summer of 2005.
O.~D. and J.~R. are grateful to the MPIPKS  Dresden 
for hospitality during 
the International Workshop
on
Collective quantum states in low-dimensional transition metal oxides
(February 22 - 25, 2005).
The authors thank 
A.~Honecker,
R.~Moessner,
K.~Penc, 
and
J.~Schulenburg
for discussions.

\end{document}